\documentclass{nature}

\title{Quantifying Information Flow During Emergencies.}

\author{Liang Gao$^{1,2}$, Chaoming Song$^3$, Ziyou Gao$^1$, Albert-L\'{a}szl\'{o} Barab\'{a}si$^{2,4,5}$,
James~P.~Bagrow$^{6,7}$ \& Dashun Wang$^{8,*}$}

\begin{document}

\maketitle

\begin{affiliations}
 \item Systems Science Institute and MOE Key Laboratory for Urban
Transportation Complex Systems Theory and Technology, Beijing Jiaotong
University, Beijing 100044, China.
 \item Center for Complex Network Research, Department of Physics,
Northeastern University, Boston, MA 02115, USA.
 \item Department of Physics, University of Miami, Coral Gables, FL 33146, USA.
 \item Center for Cancer Systems Biology, Dana-Farber Cancer Institute,
Boston, MA 02115, USA.
 \item Department of Medicine, Brigham and Women's Hospital, Harvard Medical
School, Boston, MA 02115, USA.
 \item Department of Engineering Sciences and Applied Mathematics, Northwestern Institute on Complex Systems, Northwestern University, Evanston, IL 60208, USA.
 \item Department of Mathematics and Statistics, Vermont Advanced Computing Center, Complex Systems Center, University of Vermont, Burlington, VT 05405, USA.
 \item IBM Thomas J. Watson Research Center, Yorktown Heights, New York 10598, USA.
 \item[*] Correspondence and requests for materials should be addressed to D.W.~(email: dashun@us.ibm.com).
\end{affiliations}

\noindent

\begin{abstract}
    Recent advances on human dynamics have focused on the normal patterns of
    human activities, with the quantitative understanding of human behavior
    under extreme events remaining a crucial missing chapter. This has a wide array of
    potential applications, ranging from emergency response and detection to traffic
    control and management. Previous studies have shown that human communications
    are both temporally and spatially localized following the onset of emergencies,
    indicating that social propagation is a primary means to propagate
    situational awareness. We study real anomalous events using country-wide mobile
    phone data, finding that information flow during emergencies is dominated
    by repeated communications. We further demonstrate that the observed
    communication patterns cannot be explained by inherent reciprocity in
    social networks, and are universal across different demographics.
\end{abstract}

Much effort has been devoted to the study of human dynamics under regular
and stationary situations~\cite{Eubank2004,Hufnagel2004,Brockmann2006,Colizza2006,Wu2007,Onnela2007,Gonzalez2008,Goncalves2008,Vespignani2009,Gabrielli2009,Rybski2009,Singer2009,Lazer2009,Song2010,Ratkiewicz2010,Rybski2012}.
Our quantitative
understanding of human behavior under extreme conditions, such as violent
conflicts~\cite{Bohorquez2009}, life-threatening epidemic outbreaks
\cite{Eubank2004,Colizza2006,Balcan2009,Meloni2011}, and other large-scale
emergencies, remains limited however. Yet, it is essential for a number of
practical problems faced by emergency response~\cite{Petrescu-PrahovaM2008}.
There is an extraordinary need, therefore, to quantitatively study human
dynamics and social interactions under rapidly changing or unfamiliar conditions.

Previous studies~\cite{Bagrow2011,Lu2012a} have suggested that mobile phones
can act as \emph{in situ} sensors for human behavior during anomalous events,
finding that the occurrence of anomalous events triggers a large spike in the
communication activity of those who witnessed the event. More specifically, they found that
communication spikes following emergencies are temporally and spatially
localized, indicating information flow through the social networks of affected
individuals becomes an important means to spread situational awareness and
information to the general population.

In this work, we quantify the propagation of real-world emergency information
through the contact networks of mobile phone users.
We denote the group of users directly affected by an emergency by population $G_0$,
while users they contact during the timespan of the emergency that are not
in $G_0$ form the population $G_1$.
We focus here on how $G_1$ users change their communication pattern following an
emergency. To be specific, we address three questions: First, to what magnitude
do $G_1$ users change their communication behavior? Do they show the same volume
spike as previously observed for $G_0$ users~\cite{Bagrow2011}? Second, what
is the origin of the behavior changes of $G_1$ users? What is the dominant
feature of these changes? Third, will a $G_1$ user prefer to \emph{call back}
the $G_0$ user, potentially offering comfort, support and seeking pertinent information?
Or will they instead \emph{call
forward} to propagate their situational awareness to others.
Intuitively we expect that $G_1$ users were chosen for contact by the $G_0$
users due to important relationship(s) between them, and they may communicate with each
other more often than with other peers even during normal days. It is therefore
important for future emergency detection and intervention to know whether or not there
is abnormal reciprocal communication during emergencies, compared with
ordinary activity levels.


\section*{Data and Events}
\label{sec:DataEvents}

In this paper, we use a de-identified dataset from a large mobile phone company
in a European country~\cite{Barabasi2005,Oliveira2005,Vazquez2006,Onnela2007,Goh2008,Candia2008,Gonzalez2008,Wang2009,Wang2009a,Song2010,Song2010a,Wang2011,Karsai2012,Simini2012,Jo2012,Bagrow2013}. The data consist
of approximately 10 million users and four years of cell phone activity, including
both voice calls and text messages. Each data entry records the user initiating the
call or text (caller) and the user receiving it (callee); the cellular tower
that routed the call; and the date and time when it occurred. The locations
(longitude and latitude) of cellular towers are also recorded, allowing us
to infer the location of callers whenever they initiate a communication. Hence, given the
spatiotemporal localization of an event, these data offers a unique opportunity
to quantify the social response of the affected population.

To study real events covered by this mobile phone data, we need to determine their
time and locations. We study the event set identified in previous studies~\cite{Bagrow2011},
where the authors
used Google local news (\verb"news.google.com") service to search for news
stories covering the country and time period of the mobile phone dataset.
Keywords such as `emergency', `disaster', `concert', etc.\ were used to find
potential news stories. Important events such as bombings, earthquakes and
concerts are prominently covered in the social media. Study
of these reports typically gave the precise time and the locations for these
events~\cite{Bagrow2011}.

To identify the beginning and the end of an event, $t_\mathrm{start}$ and $t_\mathrm{stop}$, we
adopt the following procedure~\cite{Bagrow2011}. First, we scan all calls in the
event region during the day covering the event, giving the event day call volume
time series (number of calls per minute) $V_\mathrm{event}(t)$~\cite{Bagrow2011}. Then, we scan calls for a
number of ``normal'' days, those modulo one week from the event day, exploiting
the weekly periodicity of $V(t)$. These normal days' call volume time
series are averaged to get $\langle V_\mathrm{normal} \rangle$. To smooth the time series,
call volumes were binned into $10$ minute intervals. The standard deviation
$\sigma(V_\mathrm{normal})$ as a function of time is then used to compute $z(t) = \Delta
V(t)/\sigma(V_\mathrm{normal})$, where $\Delta V(t) = V_\mathrm{event}(t) - \langle V_\mathrm{normal} \rangle$
is the call volume change during the event day. Finally, the interval $[t_\mathrm{start},
t_\mathrm{stop}]$ was the longest contiguous run of time intervals where $z(t) >
z_{thr}$, for some fixed cutoff $z_{thr}$. To be consistent with pervious
studies~\cite{Bagrow2011}, we chose $z_{thr} = 1.5$ for all events.

\section*{Results}
\label{sec:results}

To extract the contact network between users during an event, we track all
outgoing calls in order of occurrence during the event's time interval
$[t_\mathrm{start}, t_\mathrm{stop}]$. We therefore identified the individuals located within
the event region ($G_{0}$), as well as a $G_{1}$ group consisting of individuals
outside the event region but who receive calls from the $G_{0}$ group during the
event, a $G_{2}$ group that receive calls from $G_{1}$, and so on.

To determine how unusual the observed activities are, we compare the
call volume during the event to the average call volumes of a number of
``normal'' days (Sec.~\ref{sec:DataEvents}). Since a temporal contact network
can always be constructed from mobile phone dataset, even when no event occurs,
it is  necessary  to design a proper control for normal days to make the call
volumes between event day and normal days comparable.

To design a proper control, we study new ``cascades'' generated by the same
eyewitness users $G_{0}$ during the same time interval $[t_\mathrm{start}, t_\mathrm{stop}]$
for each normal day, and create a different cascade $\{G_{0}, g_{1}, g_{2},\cdots\}$
for each normal day.
\textbf{$G_i$ are for event period whereas $g_i$ are for normal period. $G_i$ and $g_i$
 share the same $G_0$, that is by definition $G_0 = g_0$. e.g.
$g_1$ users are the users who receive cell-phone
communication directly from $G_0$ users, $g_i$ users ($i > 1$) are the users
who receive cell-phone communication directly from $g_{i-1}$ users.
}

The number of $G_i$ users will typically be larger than
that of $g_i$ users and $G_i$ users may be more active than $g_i$ users, so
the normal day's call volume time series, $V(t|g_i)$, must be rescaled when
compared to the event day's call volume time series, $V(t|G_i)$. For this purpose,
we multiply $V(t|g_i)$ by a constant scaling factor $a_i$,
\begin{equation}\label{eq:ai}
    a_i = {\int_{\delta t}{V(\tau|G_i)d\tau} }\Bigg/{\int_{\delta t}{V(\tau|g_i)d\tau}},
\end{equation}
where both integrals run over the same ``calibration interval'' $\delta t$, and
$\tau=0$ is the start of the selection window. For most events, we integrate
over a 24-hour period two days before the event window. The factor $a_i$ was
chosen such that the total number of calls during normal time periods for
$V(t|G_i)$ is approximately equal to $a_{i}V(t|g_i)$, equalizing the normal-day
time series and removing any biases due to $|G_i|\neq |g_i|$. This control procedure
allows us to investigate call volume patterns of different user groups.

To explore the call volume patterns in different user groups, we measure the call
volume change
\begin{equation}\label{eq:DV}
    \Delta V(t|G_i)=V(t|G_i)-\langle a_{i}V(t|g_i)\rangle
\end{equation}
for $G_0$ and $G_1$ groups as a function of time, where $V(t|G_i)$ is the call
volume in the event day and $\langle a_{i}V(t|g_i)\rangle$ is the call volume
averaged over selected ``normal'' days.

Previous work has studied several aspects of communication patterns, and
found a spike in the volume of phone call activity during an emergency
event~\cite{Bagrow2011}.
Yet, by using the control mentioned above, we find
the call volume change for different groups such as $G_0$ and $G_1$ exhibits
different patterns in different events (Fig.~\ref{fig:VGi}).
More specifically, we observe activity spikes in both $G_0$ and $G_1$ groups for three
emergency events, referred to as ``Jet Scare'', ``Plane Crash'' and ``Bombing'' (Fig.~\ref{fig:VGi}{a-c}).
Yet in all other events, there is no volume spike for the $G_1$ group, e.g.
``Concert'' (Fig.~\ref{fig:VGi}{d}). These results are consistent
with previous findings~\cite{Bagrow2011}, showing that the users in the $G_1$ group are
triggered to a higher communication level, characterized by a sharp increase in call
volumes, during the emergency events. Yet, it is somewhat puzzling
that the call volume change of $G_{1}$ users have a spike, which is
instantaneous and shows virtually no delay to the spike of $G_0$ users'.
\textbf{Note the spikes of $G_0$ and $G_1$ users are synchronous qualitatively,
and sensitive to the time aggregation (see Supplementary Materials).}
As the activity spikes of $G_0$ users for emergency events are both
temporally and spatially localized, the communication of $G_1$ users
becomes the most important means of spreading situational awareness.

To quantify the reach of situational awareness, we focus on $G_1$ and study
their communication patterns after receiving a phone call or message from $G_0$.
As an example, we choose three $G_0$ users (diamonds) and
their related $G_1$ users (circles) and $G_2$ users (triangles)
in the Bombing event as a sample contact network (Fig.~\ref{fig:illus}{a}).
There are three types of communication behaviors for users in $G_1$:
(1) call back to $G_0$ user{(s)}, which are edges in orange, (2) call forward
to $G_2$ users, edges in purple, and (3) calls to other $G_1$ user{(s)}, edges
in green. We denote these three kinds of communication behaviors with,
$C_{10}$, $C_{12}$, and $C_{11}$, respectively. We measure the
contributions of these three communication modes to the total activities of $G_1$
users, finding $C_{11}$ constitutes no more than $5\%$ of the total volume of $G_1$
users' communicating activities (Fig.~\ref{fig:illus}{b}).
The observed low volume of $C_{11}$ among $G_{1}$ users during
emergencies is somewhat unexpected, given the importance of triadic closure in social
communications \cite{Granovetter1973,Borgatti2009}.
Hence, the spike
observed in $G_{1}$ users in Fig.~\ref{fig:VGi} is mainly determined by $C_{10}$
and $C_{12}$.

The existence of different communication modes in $G_1$ (Fig.~\ref{fig:illus}{a})
raises an important question: what is the temporal contribution of $C_{10}$
and $C_{12}$ to the observed spikes in $G_1$ users' activities? To this end,
we decompose $\Delta V(t|G_1)$ into $V(C_{10})$ and $V(C_{12})$ by modifying the
rescaling framework
\begin{equation}\label{eq:C1i}
    V(C_{1i})=V_{1i}(t|G_1)-\langle a_{1i}V_{1i}(t|g_1)\rangle
\end{equation}
for $i = \{0, 2\}$, where \textbf{$V_{1i}$ is the call volume from $G_1$ ($g_1$)
users to $G_i$ ($g_i$) users, and}
$a_{1i}$ is a scaling factor modified from
Eq.~\ref{eq:ai} as
\begin{equation}\label{eq:a1i}
    a_{1i} = {\int_{\delta t}{V_{1i}(\tau|G_1)d\tau} }\Bigg/{\int_{\delta t}{V_{1i}(\tau|g_1)d\tau}}.
\end{equation}

In Fig.~\ref{fig:VG1i}, we show $V(C_{10})$ and $V(C_{12})$ as a function of
time. We find that in all three emergency events, $V(C_{10})$ has evident volume
spikes during the event period. And, by comparing Fig.~\ref{fig:VGi} and
Fig.~\ref{fig:VG1i}, we find the peaks of $V(C_{10})$ are in close temporal
vicinity to those of $\Delta V(t|G_1)$. Yet, for Concert (Fig.~\ref{fig:VG1i}{d}),
we observed no clear volume spikes. Overall, Fig.~\ref{fig:VG1i} demonstrates
that $C_{10}$, as a major contribution to the observed spikes in $G_1$, indicating
the call back actions ($C_{10}$) in emergencies contribute more to the spike
of $\Delta V(t|G_{1})$ than the call forward actions ($C_{12}$). That is,
$G_1$ users prefer to interact back with $G_0$ users rather than contacting with new users
($G_2$), a phenomenon that limits the spreading of information. Indeed, $C_{10}$ measures the
reciprocal communications from $G_1$ to $G_0$, representing correspondence and
coordination calls between social neighbors. $C_{12}$, on the other hand, measures
the dissemination of situational awareness, corresponding to information cascades
that penetrate the underlying social network. Hence, the results in
Fig.~\ref{fig:VG1i} indicate that during emergencies both dissemination and
call-back response of emergency information are important for information flow,
and they together determine the magnitude of $G_1$ users' communication spikes
 (Fig.~\ref{fig:VGi}).

These preceding results indicate that reciprocal communications play a dominant role in
social response during emergency, raising an important question: what is the
origin of the observed reciprocal correspondence? There are two possible
mechanisms likely at work here. First is the heterogeneous nature of
reciprocities: Social ties are characterized by different reciprocities
\cite{Newman2002,Skvoretz2007,Hidalgo2008,Kovanen2010,Wang2011a,Doorn2012}, corresponding
to different likelihood of reciprocal communications upon receiving a call from others.
Therefore, the large increase in reciprocal communications may represent a
selection bias introduced by eyewitnesses by communicating with their social
neighbors with high reciprocity. The second possible factor is a behavioral change
of social neighbors after learning about the event, corresponding to coordination
and providing additional information to eyewitnesses. To quantify the competition
between these two factors, we measure the reciprocity of communications between
any two individuals, during normal periods.

In an unweighted network, a general definition of link reciprocity $R$ is to
measure the tendency of two nodes to form mutual connections
($A\rightarrow B$ and $B\rightarrow A$)~\cite{Newman2002,Garlaschelli2004,Zamora-Lopez2008}. Hence, $R=1$ for a
purely bidirectional link, and $R=0$ for a unidirectional one. Considering the
weighted nature of contact networks\cite{Kovanen2010,Wang2011a,Gavalda2012}, we
define the reciprocity of communications between $G_0$ and $G_1$ users as
\begin{equation}
    \mathbf{R = 1- \left|1- \frac{2 V_{i\rightarrow j}}{V_{i\rightarrow j}+V_{j\rightarrow i}}\right|},
    \label{eq:recip}
\end{equation}
where $V_{x\rightarrow y}$ is the number of calls from user $x$ to user $y$ within
a given period, and $|\cdot |$ is the absolute value. With this definition, two
users have a reciprocity ranging from $0$ to $\mathbf{1}$, where $R=\mathbf{1}$ corresponds to
reciprocal links, and $R=0$ for non-reciprocal ones.

In Fig.~\ref{fig:recip}, we show the reciprocity of communications averaged over
all pairs of users in $G_0$ and $G_1$ for the four events during event periods.
We find that, for Bombing, Plane Crash and Jet Scare, the average reciprocity
for each emergency event shows a significant increase, deviating by approximately
$1.8$, $3$ and $6$ standard deviations from normal days, respectively. This result
indicates that the observed increase in ``call-back'' actions from $G_1$ to $G_0$
during these emergency events correspond to behavior change in communications.
If the increased call-back were entirely random, the distribution of  reciprocity
over the $G_1$ population would be sufficient to explain the resulting call-back,
but we do not observe this. For Concert, we observe a decrease in reciprocity comparing to
normal periods, with $4.7\sigma$ below the averaged reciprocity of normal days,
indicating clear distinctions between emergency and non-emergency events.

To test the consistency of our results, we also study reciprocity for other
emergency and non-emergency events. As shown in Supplementary Materials Fig.~S1,
the reciprocities for Blackout and Earthquake are characterized by only modest
increase, well within the range of one standard deviation, reassuring the preceding
results on correlations between activity spikes in $G_1$ (Fig.~3a in Ref.~\cite{Bagrow2011}) and their increase in
reciprocity.

Finally, to better understand the origin of the observed increase of reciprocity, we
measure the contributions to $G_1$ users' behavioral change for different
demographics. More specifically, we obtained self-reported gender information,
available for $88\%$ of users, and consider all communications in which
gender information is available for both parties. There are four kinds of coupled
pairs between $G_0$ and $G_1$ users: male-male (MM), male-female (MF), female-male
(FM) and female-female (FF). For each event, we compute the reciprocities for
the four different kinds of pairs separately. We also average the reciprocities
of the MF and FM pairs as cross-gender pairs (CG), and the MM and FF pairs as
same-gender pairs (SG). Interestingly, we find the collective response from
different demographics is almost universal (Fig.~\ref{fig:gender}). That is,
for emergency events with significant increases in reciprocity (Jet Scare and
Plane Crash), the reciprocities across different gender pairs are all several
standard deviations larger than normal periods (Fig.~\ref{fig:recip} (a) and (b)).

\section*{Discussion}
\label{sec:Future}

Taken together, we have studied the cell phone communications during anomalous
events, and find volume spikes in $G_1$'s communication compared to normal days
in three emergencies. To uncover the possible origin of the volume spikes, we
decomposed $G_1$'s communications into call-back ($C_{10}$) and call-forward
($C_{12}$) actions. Comparing to non-emergency events, we found that the dominant
component of volume spikes is $C_{10}$ for all three emergency events, indicating
the need for correspondence with eyewitnesses is more critical than the dissemination
of situational awareness during emergencies. We further demonstrated such communication
patterns correspond to a behavior change in $G_1$ users that cannot be explained
by reciprocity or demographics. We believe the empirical findings reported in
this paper present relevant information that can be used to benchmark potential
models, and will play an increasingly important role as large-scale data flourish
and our quantitative understanding human behavior deepens.


\begin{thebibliography}{1}
\bibitem{Bagrow2013}
Bagrow, J.~P., \& Brockmann, D.
\newblock Natural emergence of clusters and bursts in network evolution.
\newblock {\em Phys. Rev. X} \textbf{3}, 021016 (2013).

\bibitem{Bagrow2011}
Bagrow, J.~P., Wang, D., \& Barab\'{a}si, A.-L.
\newblock Collective response of human populations to large-scale emergencies.
\newblock {\em PLoS ONE} \textbf{6}, e17680 (2011).

\bibitem{Balcan2009}
Balcan, D. et al.
\newblock Multiscale mobility networks and the spatial spreading of infectious
  diseases.
\newblock {\em Proc. Natl. Acad. Sci. USA}
  \textbf{106}, 21484-21489 (2009).

\bibitem{Barabasi2005}
Barab\'{a}si, A.-L.
\newblock The origin of bursts and heavy tails in human dynamics.
\newblock {\em Nature} \textbf{435}, 207-211 (2005).

\bibitem{Bohorquez2009}
Bohorquez, J.~C., Gourley, S., Dixon, A.~R., Spagat, M., \& Johnson, N.~F.
\newblock Common ecology quantifies human insurgency.
\newblock {\em Nature} \textbf{462}, 911-914 (2009).

\bibitem{Borgatti2009}
Borgatti, S.~P., Mehra, A., Brass, D.~J., \& Labianca, G.
\newblock Network analysis in the social sciences.
\newblock {\em Science} \textbf{323}, 892-895 (2009).

\bibitem{Brockmann2006}
Brockmann, D., Hufnagel, L., \& Geisel, T.
\newblock The scaling laws of human travel.
\newblock {\em Nature} \textbf{439}, 462-465 (2006).

\bibitem{Candia2008}
Candia, J. et al.
\newblock Uncovering individual and collective human dynamics from mobile phone
  records.
\newblock {\em J. Phys. A} \textbf{41}, 224015 (2008).

\bibitem{Colizza2006}
Colizza, V., Barrat, A., Barth\'{e}lemy, M., \& Vespignani, A.
\newblock The role of the airline transportation network in the prediction and
  predictability of global epidemics.
\newblock {\em Proc. Natl. Acad. Sci. USA}
  \textbf{103}, 2015-2020 (2006).

\bibitem{Eubank2004}
Eubank, S. et al.
\newblock Modelling disease outbreaks in realistic urban social networks.
\newblock {\em Nature} \textbf{429}, 180-184 (2004).

\bibitem{Gabrielli2009}
Gabrielli, A. \& Caldarelli, G.
\newblock Invasion percolation and the time scaling behavior of a queuing model
  of human dynamics.
\newblock {\em J. Stat. Mech. Theor. Exp.}, P02046 (2009).

\bibitem{Garlaschelli2004}
Garlaschelli, D. \& Loffredo, M.~I.
\newblock Patterns of link reciprocity in directed networks.
\newblock {\em Phys. Rev. Lett.} \textbf{93}, 268701 (2004).

\bibitem{Gavalda2012}
Gavalda, A., Duch, J., \& G\'omez-Garde\~nes, J.
\newblock Reciprocal interactions out of congestion-free adaptive networks.
\newblock {\em Phys. Rev. E} \textbf{85}, 026112 (2012).

\bibitem{Goh2008}
Goh, K.-I. \& Barab\'{a}si, A.-L.
\newblock Burstiness and memory in complex systems.
\newblock {\em EPL (Europhys. Lett)} \textbf{81}, 48002 (2008).

\bibitem{Goncalves2008}
Goncalves, B. \& Ramasco, J.~J.
\newblock Human dynamics revealed through web analytics.
\newblock {\em Phys. Rev. E} \textbf{78}, 026123 (2008).

\bibitem{Gonzalez2008}
Gonz\'{a}lez, M.~C., Hidalgo, C.~A., \& Barab\'{a}si, A.-L.
\newblock Understanding individual human mobility patterns.
\newblock {\em Nature} \textbf{453}, 779-782 (2008).

\bibitem{Granovetter1973}
Granovetter, M.~S.
\newblock The strength of weak ties.
\newblock {\em Am. J. Sociol.} \textbf{78}, 1360-1380 (1973).

\bibitem{Hidalgo2008}
Hidalgo, C.~A. \& Rodriguez-Sickert, C.
\newblock The dynamics of a mobile phone network.
\newblock {\em Physica A} \textbf{387}, 3017-3024 (2008).

\bibitem{Hufnagel2004}
, L., Brockmann, D., \& Geisel, T.
\newblock Forecast and control of epidemics in a globalized world.
\newblock {\em Proc. Natl. Acad. Sci. USA}
  \textbf{101}, 15124-15129 (2004).

\bibitem{Jo2012}
Jo, H.-H., Karsai, M., Kert\'{e}sz, J., \& Kaski, K..
\newblock Circadian pattern and burstiness in mobile phone communication.
\newblock {\em New J. Phys.} \textbf{14}, 013055 (2012).

\bibitem{Karsai2012}
Karsai, M., Kaski, K., Barab\'{a}si, A.-L., \& Kert\'{e}sz, J.
\newblock Universal features of correlated bursty behaviour.
\newblock {\em Sci. Rep.} \textbf{2}, 397 (2012).

\bibitem{Kovanen2010}
Kovanen, L., Saramaki, J., \& Kaski, K.
\newblock Reciprocity of mobile phone calls.
\newblock {\em Dynamics of Socio-Economic Systems} \textbf{2}, 138-151 (2011).

\bibitem{Lazer2009}
Lazer, D. et al.
\newblock Social science: Computational social science.
\newblock {\em Science} \textbf{323}, 721-723 (2009).

\bibitem{Lu2012a}
Lu, X., Bengtsson, L., \& Holme, P.
\newblock Predictability of population displacement after the 2010 haiti
  earthquake.
\newblock {\em Proc. Natl. Acad. Sci. USA}
  \textbf{109}, 11576-11581 (2012).

\bibitem{Meloni2011}
Meloni, S. et al.
\newblock Modeling human mobility responses to the large-scale spreading of
  infectious diseases.
\newblock {\em Sci. Rep.} \textbf{1}, 62 (2011).

\bibitem{Newman2002}
Newman, M.~E.~J., Forrest, S., \& Balthrop, J.
\newblock Email networks and the spread of computer viruses.
\newblock {\em Phys. Rev. E} \textbf{66}, 035101 (2002).

\bibitem{Oliveira2005}
Oliveira, J.~G. \& Barab\'{a}si, A.-L.
\newblock Human dynamics: Darwin and einstein correspondence patterns.
\newblock {\em Nature} \textbf{437}, 1251-1251 (2005).

\bibitem{Onnela2007}
Onnela, J.-P. et al.
\newblock Structure and tie strengths in mobile communication networks.
\newblock {\em Proc. Natl. Acad. Sci. USA}
  \textbf{104}, 7332-7336 (2007).

\bibitem{Petrescu-PrahovaM2008}
Petrescu-Prahova, M. \& Butts, C.
\newblock Emergent coordinators in the world trade center disaster.
\newblock {\em International Journal of Mass Emergencies and Disasters}
  \textbf{26}, 133–168 (2008).

\bibitem{Ratkiewicz2010}
Ratkiewicz, J., Fortunato, S., Flammini, A., Menczer, F., \& Vespignani, A.
\newblock Characterizing and modeling the dynamics of online popularity.
\newblock {\em Phys. Rev. Lett.} \textbf{105}, 158701 (2010).

\bibitem{Rybski2009}
Rybski, D., Buldyrev, S.~V., Havlin, S., Liljeros, F., \& Makse, H.~A.
\newblock Scaling laws of human interaction activity.
\newblock {\em Proc. Natl. Acad. Sci. USA}
  \textbf{106}, 12640-12645 (2009).

\bibitem{Rybski2012}
Rybski, D., Buldyrev, S.~V., Havlin, S., Liljeros, F., \& Makse, H.~A.
\newblock Communication activity in a social network: relation between
  long-term correlations and inter-event clustering.
\newblock {\em Sci. Rep.} \textbf{2}, 560 (2012).

\bibitem{Simini2012}
Simini, F., Gonz\'{a}lez, M.~C., Maritan, A., \& Barab\'asi, A.-L.
\newblock A universal model for mobility and migration patterns.
\newblock {\em Nature} \textbf{484}, 96-100 (2012).

\bibitem{Singer2009}
Singer, H.~M., Singer, I., \& Herrmann, H.~J.
\newblock Agent-based model for friendship in social networks.
\newblock {\em Physical Review E} \textbf{80}, 026113 (2009).

\bibitem{Skvoretz2007}
Skvoretz, J. \& Agneessens, F.
\newblock Reciprocity, multiplexity, and exchange: Measures.
\newblock {\em Quality \& Quantity} \textbf{41}, 341-357 (2007).

\bibitem{Song2010a}
Song, C., Koren, T., Wang, P., \& Barab\'asi, A.-L.
\newblock Modelling the scaling properties of human mobility.
\newblock {\em Nat Phys} \textbf{6}, 818–823 (2010).

\bibitem{Song2010}
Song, C., Qu, Z., Blumm, N., and Barab\'{a}si, A.-L.
\newblock Limits of predictability in human mobility.
\newblock {\em Science} \textbf{327}, 1018-1021 (2010).

\bibitem{Doorn2012}
G.~S. van Doorn \& M.~Taborsky.
\newblock The evolution of generalized reciprocity on social interaction
  networks.
\newblock {\em Evolution} \textbf{66}, 651-664 (2012).

\bibitem{Vazquez2006}
Vazquez, A. et al.
\newblock Modeling bursts and heavy tails in human dynamics.
\newblock {\em Physical Review E} \textbf{73}, 036127 (2006).

\bibitem{Vespignani2009}
Vespignani A.
\newblock Predicting the behavior of techno-social systems.
\newblock {\em Science} \textbf{325}, 425-428 (2009).

\bibitem{Wang2011a}
Wang, C. et al.
\newblock Weighted reciprocity in human communication networks.
\newblock arXiv:1108.2822 (2011).

\bibitem{Wang2011}
Wang, D., Pedreschi, D., Song, C., Giannotti, F., \& Barab\'{a}si, A.-L.
\newblock Human mobility, social ties, and link prediction.
\newblock In {\em Proceedings of the 17th ACM SIGKDD International Conference
  on Knowledge Discovery and Data Mining} \textbf{KDD'11}, 1100-1108, New York,
  NY, USA, (2011).

\bibitem{Wang2009}
Wang, P. \& Gonz\'{a}lez, M.~C.
\newblock Understanding spatial connectivity of individuals with non-uniform
  population density.
\newblock {\em Phil. Trans. R. Soc. A} \textbf{367}, 3321-3329 (2009).

\bibitem{Wang2009a}
Wang, P., Gonz\'alez, M.~C., Hidalgo, C.~A., \& Barab\'asi, A.-L.
\newblock Understanding the spreading patterns of mobile phone viruses.
\newblock {\em Science} \textbf{324}, 1071-1076 (2009).

\bibitem{Wu2007}
Wu, F. \& Huberman, B.~A.
\newblock Novelty and collective attention.
\newblock {\em Proc. Natl. Acad. Sci. USA}
  \textbf{104}, 17599-17601 (2007).

\bibitem{Zamora-Lopez2008}
Zamora-L\'opez, G., Zlati\ifmmode~\acute{c}\else \'{c}\fi{}, V., Zhou, C.,
  \ifmmode \check{S}\else \v{S}\fi{}tefan\ifmmode \check{c}\else
  \v{c}\fi{}i\ifmmode~\acute{c}\else \'{c}\fi{}, H., \& Kurths, J.
\newblock Reciprocity of networks with degree correlations and arbitrary degree
  sequences.
\newblock {\em Phys. Rev. E} \textbf{77}, 016106 (2008).
\end{thebibliography}


\section*{Acknowledgements}
This work was supported by the Network Science Collaborative Technology Alliance
sponsored by the US Army Research Laboratory under Agreement Number
W911NF-09-{2}-0053; the Office of Naval Research under Agreement Number
N000141010968; the Defense Threat Reduction Agency awards WMD BRBAA07-J-{2}-0035
and BRBAA08-Per4-C-{2}-0033; and the James S. McDonnell Foundation 21st Century
Initiative in Studying Complex Systems. L. Gao thanks the financial support from
the Major State Basic Research Development Program of China (973 Program)
No. 2012CB725400, the China Scholarship Council, the National
Natural Science Foundation of China (71101009, 71131001), and the Fundamental Research
Funds for the Central Universities No. 2012JBM067.

\section*{Author contributions}
L.G., C.S., and D.W. designed research; L.G., and D.W. performed research;
L.G., J.B., and D.W. analyzed data;
L.G., Z.G., A.B., J.B., and D.W. wrote the paper.

\section*{Additional information}
\textbf{Supplementary information} accompanies this paper at http://www.nature.com/scientificreports
\noindent
\textbf{Competing financial interests:} The authors declare no competing financial interests.

\begin{figure}
\begin{center}
\end{center}
\caption{The temporal behavior of the call volume change for three emergency
      events and one non-emergency event. \textbf{a}, Jet Scare; \textbf{b},
      Plane Crash; \textbf{c}, Bombing; \textbf{d}, Concert. Call volume changes
      for the event are compared to the average rescaled call volume change of
      five corresponding normal days to compute $\Delta V(t|G_i)$. Dark blue line is for $G_0$ users, dark
      red line is for $G_1$ users. Vertical dash line in red is the start time
      of the event.
\label{fig:VGi}
}
\end{figure}

\begin{figure}
\begin{center}
\end{center}
\caption{There are three kinds of communicating behaviors for $G_1$ users.
    \textbf{a}, A random sample contact network during the Bombing.
    The $G_0$, $G_1$, and $G_2$ users are in red (diamond), yellow (circle), and
    blue (triangle), respectively. Edges in orange, purple, and green represent
    call back ($C_{10}$), call forward ($C_{12}$), and calls to other $G_{1}$
    users ($C_{11}$), respectively. \textbf{b}, Histogram demonstratinghow
    strongly each communication pattern contributes to the total communication
    activity. Only $\approx 5\%$ of $V{(G_1)}_{event}$ is due to $V(C_{11})$
    activity for example.
\label{fig:illus}
}
\end{figure}

\begin{figure}
\begin{center}
\end{center}
\caption{The time dependence of call volume $V(C_{10})$ and $V(C_{12})$ for
      three emergency events and one non-emergency event. Solid (orange) line is call
      back volume $V(C_{10})$, dash (purple) line is call forward volume $V(C_{12})$.
      Vertical dash lines in red are the start and the end times of the event.
\label{fig:VG1i}
}
\end{figure}

\begin{figure}
\begin{center}
\end{center}
  \caption{Histogram for the average reciprocity (Eq.~\ref{eq:recip}) of communications
      during event period for three emergency events and one non-emergency event. The
      reciprocities of communications are averaged over all observed links
      between $G_0$ and $G_1$ users for the event day and each normal day.
      Event day's value is in red. The final averaged value for normal day is in
      blue, and averaged over five normal days' averaged reciprocities. The
      standard deviation of five normal days' averaged reciprocities are also
      shown as a error bar for each event.
\label{fig:recip}
}
\end{figure}

\begin{figure}
\begin{center}
\end{center}
\caption{Gender influence on reciprocity. Histogram for the average reciprocity
      of communications during three emergency events [(a) Jet Scare, (b) Plane
      Crash, (c) Bombing] and one non-emergency event [(d) Concert]. In (a) and (b),
      there is a strong change in R which is unaffected by gender. In (c) and (d),
      however, reciprocity does not generally change in a significant manner.
      Error bars denote the standard deviation.
\label{fig:gender}
}
\end{figure}

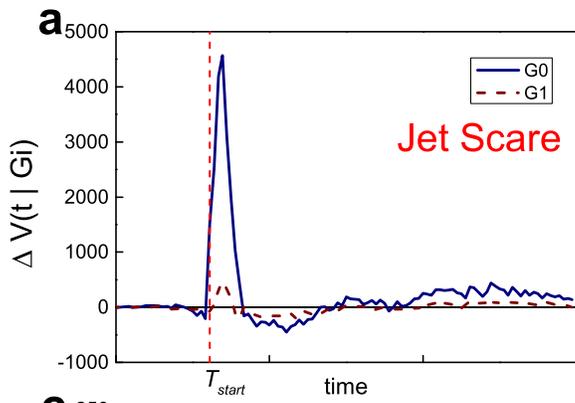
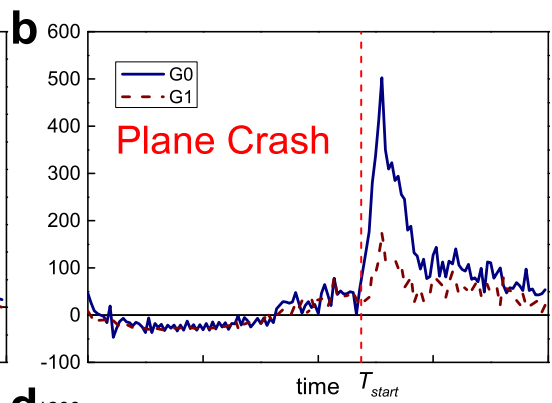
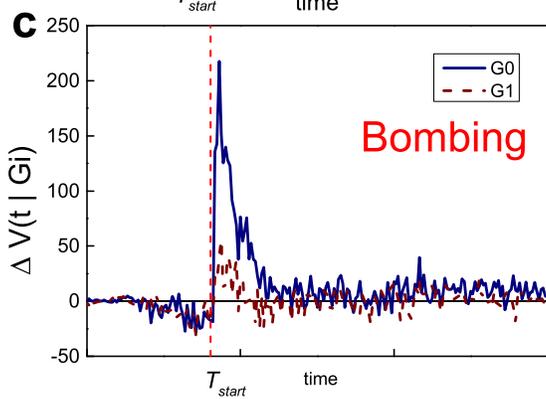
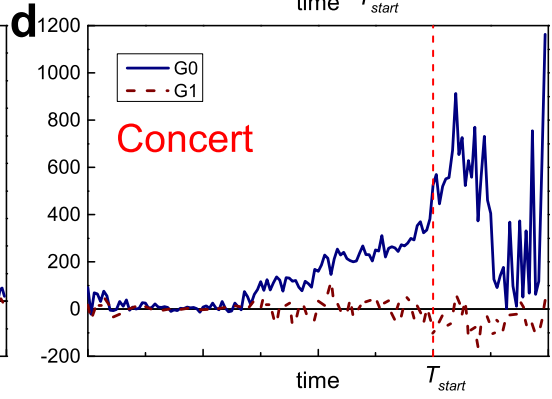
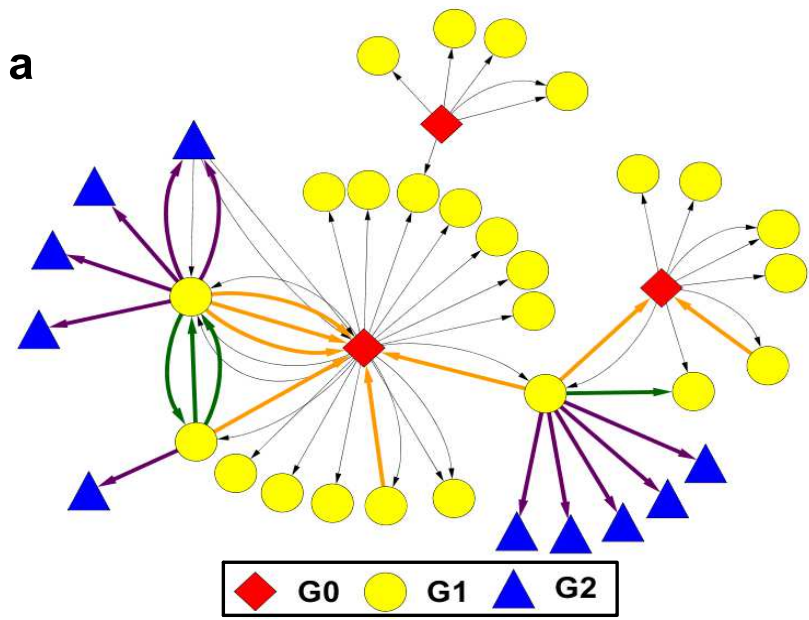
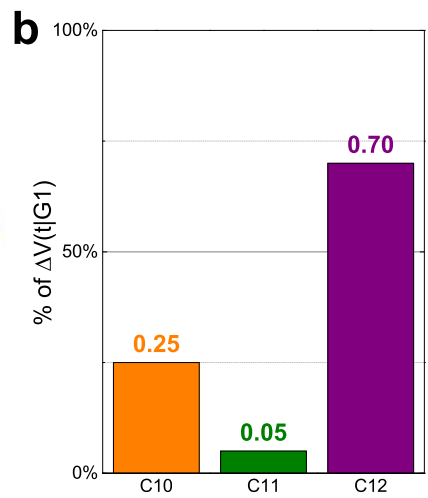
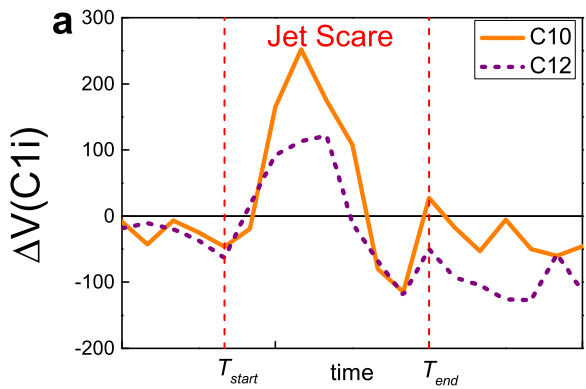
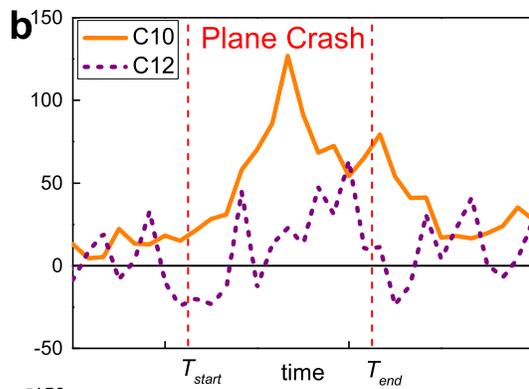
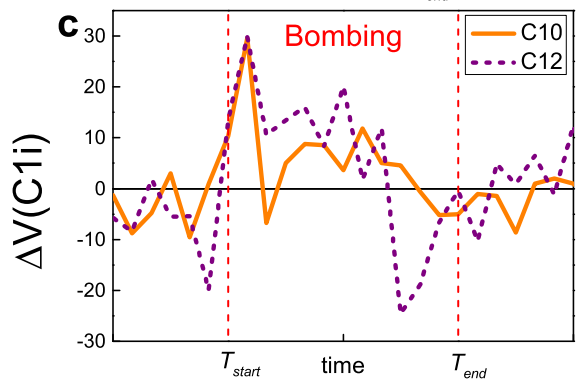
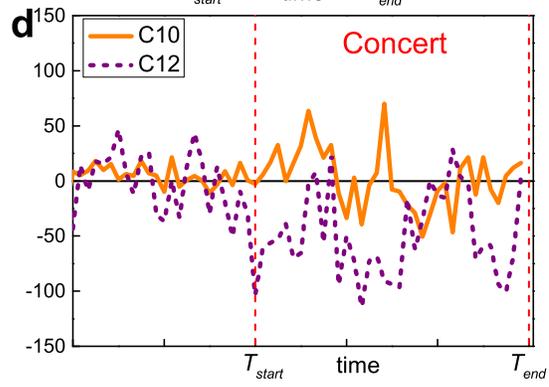
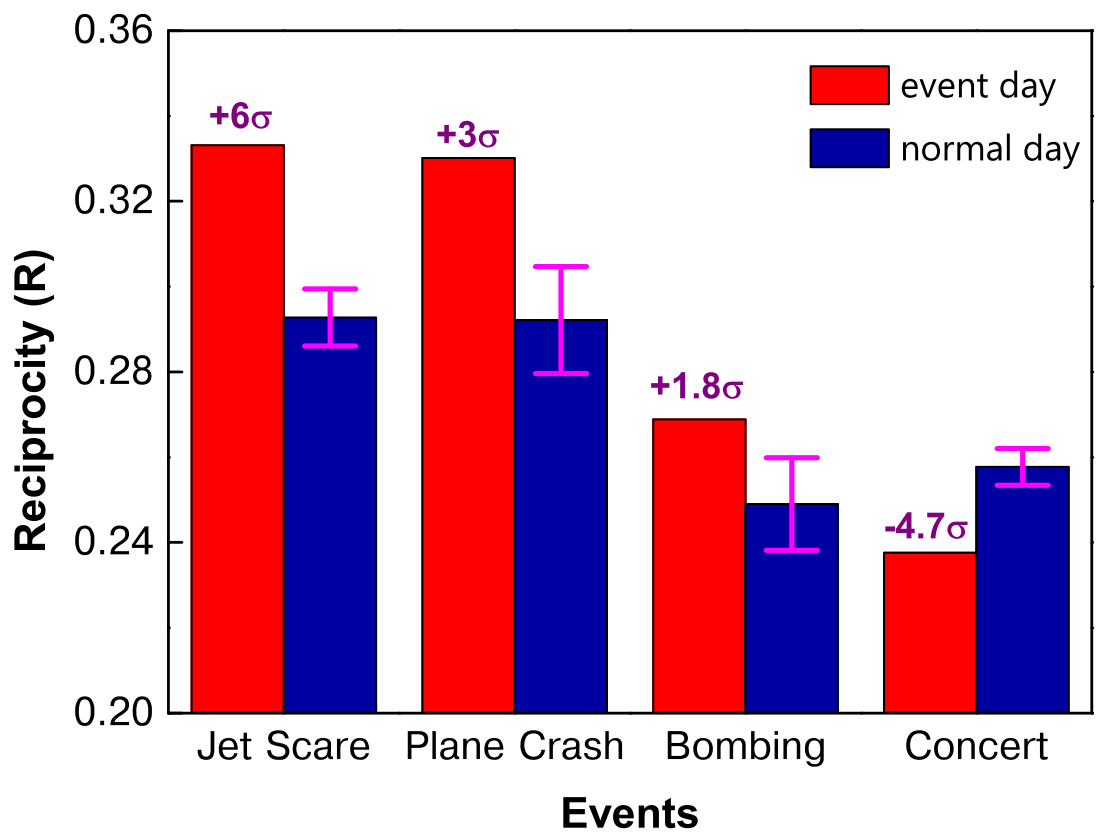
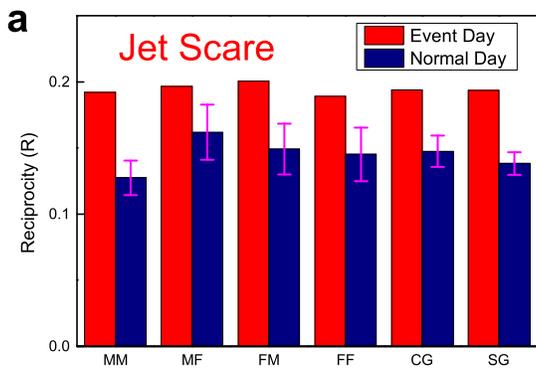
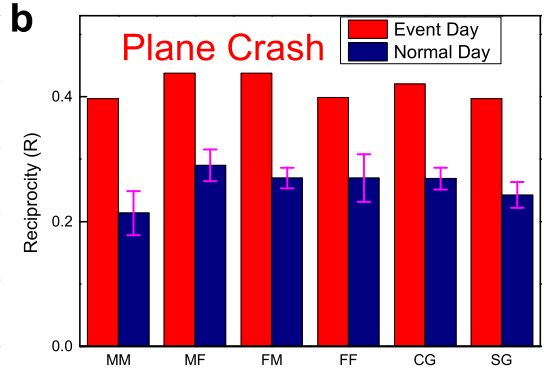
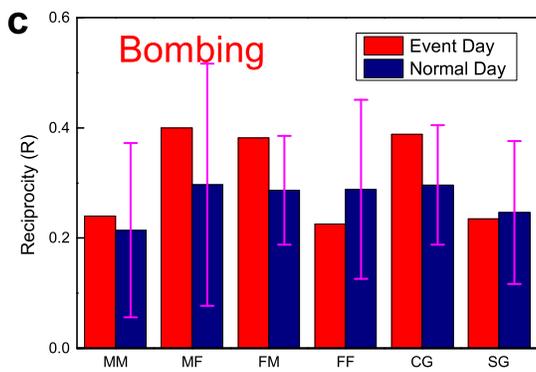
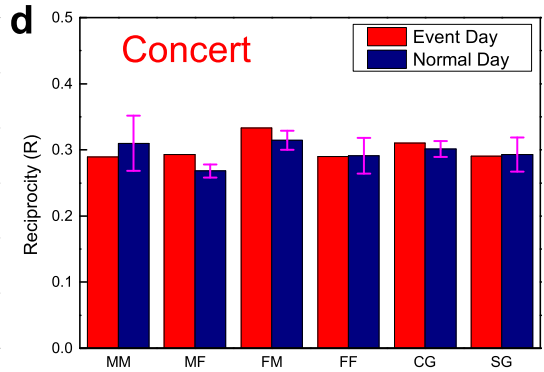
\end{document}